\documentclass[mathleft
]{an}
\usepackage{graphicx}
\usepackage{times}
\usepackage{natbib}
\bibpunct{(}{)}{;}{a}{}{}
\newcommand{\citeN}[1]{\citeauthor{#1} (\citeyear{#1})}
\newcommand{\citeNP}[1]{\citeauthor{#1} \citeyear{#1}}

\newcommand{\CaII}{{Ca}{\rm II}}
\newcommand{\HeI}{{He}{\rm I}}
\newcommand{\FeI}{{Fe}{\rm I}}

\begin{document}

\Pagespan{789}{}
\Yearpublication{2006}%
\Yearsubmission{2005}%
\Month{11}%
\Volume{999}%
\Issue{88}%

\title{Spectro-polarimetry in the era of large solar telescopes}

\author{H. Socas-Navarro\thanks{Corresponding author:
  \email{hsocas@iac.es}\newline}
}
\titlerunning{Spectro-polarimetry with large solar telescopes}
\authorrunning{H. Socas-Navarro}
\institute{
Instituto de Astrof\'\i sica de Canarias, Avda V\'\i a L\'actea S/N,
La Laguna 38200, Tenerife, Spain}

\received{30 Dec 2009}
\accepted{}
\publonline{later}


\abstract{This paper discusses some of the challenges of
  spectro-polarimetric observations with a large aperture solar
  telescope such as the ATST or the EST. The observer needs to reach a
  compromise among spatial and spectral resolution, time cadence, and
  signal-to-noise ratio, as only three of those four parameters can be
  pushed to the limit. Tunable filters and grating spectrographs
  provide a natural compromise as the former are more suitable for
  high-spatial resolution observations while the latter are a better
  choice when one needs to work with many wavelengths at full
  spectral resolution. Given the requirements for the new science
  targeted by these facilities, it is important that 1)tunable filters
  have some multi-wavelength capability; and 2)grating spectrographs
  have some 2D field of view.
}

\maketitle

\section{Introduction}

The Advanced Technology Solar Telescope (ATST), the European Solar
Telescope (EST) and the so-called plan B for the space mission
Solar-C are all being designed with multi-wavelength
spectro-polarimetry as one of the top priorities. This kind of
observations will permit novel studies of the processes taking place
through the solar atmosphere, from the photosphere to the corona, as
they traverse very different physical regimes with conditions that
typically vary over several orders of magnitude.

Observing the chromosphere/corona at disparate wavelengths with
spectro-polarimetry is still challenging even with the large-aperture
telescopes currently under development. The science goals dictate
requirements that typically include the following:
\begin{itemize}
\item High spatial resolution. We are interested in observing
  processes that take place at very small spatial scales.
  Thus, the telescope needs to be
  able to observe near the diffraction limit.
\item High cadence. The demand for high spatial resolution imposes
  also requirements on the time cadence. If a particular feature is
  moving with a projected velocity $v$, integrating over time scales
  comparable or longer than $t={\Delta x}/v$ (where $\Delta x$ is the
  spatial resolution) will result in image degradation. The sound
  speed (which is often taken as a first estimate for $v$) in the
  photosphere is typically around 6~km~s$^{-1}$. However, in the
  chromosphere it is significantly larger and one often finds
  supersonic flows.
\item High spectral resolution. Many science goals require not only
  imaging but also the observation of detailed line profiles to be
  able to track shock waves in the shape of chromospheric lines, as
  well as other interesting phenomena that produce emission features
  or other imprints on the intensity or polarization spectral
  profiles. Ideally one would want to resolve the Doppler width of the
  lines. Many interesting chromospheric lines are from H and He ions
  and have large Doppler widths, but others (particularly Ca) impose
  requirements for a better resolution.
\item High signal-to-noise ratio. Polarimetric signals are often weak,
  especially those coming from the chromosphere. Typically, a
  continuum signal-to-noise ratio of 1000 is required for
  diagnostics based on the Zeeman effect (some times an order of
  magnitude better is required for Hanle-effect studies). 
\end{itemize}

Unfortunately, it is not possible to meet all four of these
requirements simultaneously, regardless of the telescope
size. Observers will have to make compromises and optimize the
instrument configuration for the problem at hand. Compromising will
then be a key notion in observational solar physics for the next
decade. 

\section{Tunable filters vs grating spectrographs}
\label{TFGS}

The nature of instrumentation itself provides us with a first natural
set of choices. In recent years, technological advances have brought
the development of tunable filters (TFs) to a point where the narrow
passbands and tuning capabilities permit the observation of
spectrally-resolved profiles, providing similar functionalities to
those of conventional grating spectrographs (GSs). However, TFs and
GSs exhibit significant differences that one should keep in mind when
designing an observing campaign. Generally speaking, TFs make it
easier to reach the highest possible spatial resolution and are
susceptible to the application of more or less standard image
reconstruction techniques (such as speckle, phase diversity or
multi-object multi-frame blind deconvolution). In practice, one is
limited to one or two spectral lines and the interpretation of the
observed profiles may be complicated because the Sun evolves in time
during the measuring process. On the
other hand, GSs should be utilized for an optimal measurement of the
spectral properties of lines and continua, especially if this is
required at many wavelengths.

\begin{figure*}
\includegraphics[width=\textwidth]{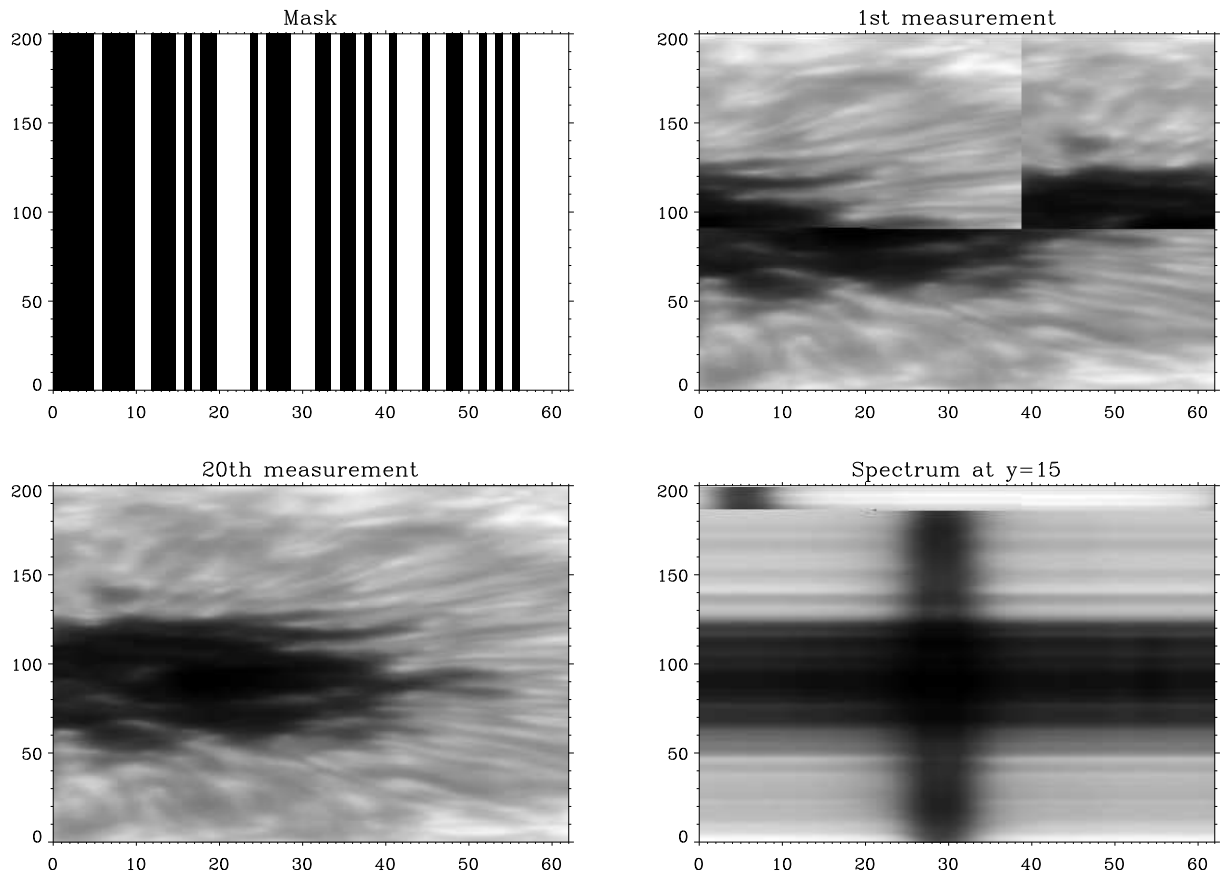}
\caption{Simulation of a TUNIS dataset using Hinode observations of a
  sunspot. Top-left: Combination of 31 slits (mask) employed to create
  the observation in the right panel. Top-right: Observed field of
  view. Each column is observed at a different combination of
  wavelengths, dictated by the mask configuration. Bottom-left:
  Another measurement with a different mask
  configuration. Bottom-right: Reconstructed spectra as if observed
  through a slit. The horizontal axis is the wavelength dispersion
  direction and the vertical is the spatial direction. The aspect
  ratio of the two ($x$,$y$) images shown corresponds to that of the
  original observation and is distorted because the slit stepping is
  different from the pixel sampling along the slit. Figure courtesy of
  A. Asensio Ramos and A. L\'opez Ariste}
\label{tunis}
\end{figure*}

Scanning a field of view with GSs is becoming increasingly
painful as the spatial resolution improves (and also the field to observe
grows with newer large-format detectors). Just to give an idea,
scanning a 50\arcsec \, field of view with diffraction-limited resolution
requires the observation of about 3000 slit positions, or some 4 hours
of observation. This is simply unacceptable. Clearly, a further step
in the evolution of GSs is necessary to at least alleviate this
issue. Several alternatives already exist with varying degrees of
maturity, namely:

\begin{itemize}
\item Multi-slit: The growing size of (typically square) detectors
 often results in pixel rows that are too large for the observation of
 the desired spectral range. As an example, with a 4k$\times$4k camera
 one could acquire a range of around 60~\AA , much larger than the free
 spectral range that one normally has. This wasteful situation may be
 avoided by using multiple slits separated just enough to avoid
 overlapping between their respective spectra (\citeNP{MRC+74};
 \citeNP{JLM+08}). Ideally, one would like 
 to have pre-filters with a square shape to facilitate multi-slit
 observations. In this manner, it should be possible to dramatically
 reduce the time required to scan a given field of view by an order of
 magnitude or even more.
\item Image slicing: This technique introduces optics in the focal
  plane to slice the field of view in the horizontal direction and
  reimage the pieces together forming a column in the slit
  direction (see, e.g., \citeNP{WKK+96}). Thus, a long slit may
  accommodate different parts of a 2D field of view.
\item Fiber optics: Here, a densely packed fiber optics bundle is used
  to again reimage a 2D portion of the focal plane onto the 1D
  spectrograph slit. Losses at the junctions of individual fibers may
  be minimized by introducing a properly designed array of
  micro-lenses (\citeNP{ASC98}).  
\item Double-pass spectrograph: A technique originally called
  subtracting double dispersion was first introduced by \citeN{S68},
  later developed by \citeN{M77} in MSDP and more recently in TUNIS
  (L\' opez Ariste {\it et al}, in preparation). With this
  technique, the light hits the grating {\it before} the field of view
  is blocked by the slit. The subsequent passage through the slit
  selects then a specific wavelength (which varies in the direction
  perpendicular to the slit) and the light is then reflected back to
  the grating. This second incidence reverses the wavelength
  dispersion and collapses the beam to form again the original 2D
  image, with the peculiarity that each column is now observed at a
  specific wavelength or combination of wavelengths. By moving the
  slit, the wavelengths observed at each column change, providing a
  simple tuning mechanism. In addition to this, it is possible to have
  multiple slits and also to reconfigure them dynamically, in order to
  scan the ($x$,$\lambda$)-space in an optimal manner to minimize the
  number of measurements required (\citeNP{ARLA10}). Fig~\ref{tunis}
  shows an example of simulated data with such a clever combination of
  slits. An interesting alternative application of this kind of
  instrument would be as a replacement for a Fabry-Perot filter for
  large-aperture telescopes, where the fabrication of a sufficiently
  large etalon could represent a technical problem.
 \end{itemize}

A more complete discussion of methods for 2D spectroscopy can be found
e.g. in \citeN{B09}. 

\section{Multi-wavelength observations}
\label{multiwavelength}

The need for observing at multiple wavelengths has become increasing
obvious in recent years for the reasons stated below. Unfortunately,
this enterprise is still very challenging. When the instrumentation
employed is not designed for this purpose, one has to coordinate the
operation of different instruments and often different telescopes as
well. Various operational and logistic issues complicate such
coordinated campaigns, including the pointing procedure to ensure that
all of the instruments are observing the same field. 

Even when the entire campaign is carried out with one telescope and
instrument, one has to deal with chromaticity issues (in the
telescope, transfer optics and instrumentation) and balancing beams
with intrinsically different photon flux (i.e., making sure that all
the detectors receive enough photons to build the necessary
signal-to-noise ratio while at the same time avoiding
saturation). This latter problem is more critical if one is also doing
polarimetry, as the modulation rate must be adequate for all of the
detectors (or, alternatively, have a separate modulator for each
beam). In that case, it is also important to ensure that the
modulation efficiency is high at all the observed wavelengths. In
general, this is not straightforward to accomplish unless one has a
monochromatic polarimeter. Finally, on top of all these difficulties,
ground-based facilities are affected by differential atmospheric
refraction. This effect, which is more pronounced in the blue side of
the spectrum, produces a shift of the image on the focal plane in the
direction connecting the horizon with the zenith. For imaging
instrumentation, such as TFs, this issue is not critical since it is
possible to correct the shift {\it a posteriori} by properly
realigning the data cubes. The same is not true for GS, however, as
the process of scanning over the field of view introduces a time lag
in the images of an object at different wavelengths. One way to
overcome this difficulty is to observe with the slit perpendicular to
the line connecting the horizon and the zenith, so that the
differential refraction shift is along the slit and then it is easier
to again make the correction {\it a posteriori} (\citeNP{F82}). In any
case, atmospheric differential refraction is another argument to push
for GSs that have at least some limited 2D capability, as discussed in
Sect~\ref{TFGS} above.

The main driver for observing simultaneously at disparate wavelengths
is of course to have information of the atmosphere at different
heights and be able to reconstruct a 3D view of the processes taking
place in it. However, there may be other motivations to do this. One
example is to have differential measurements, very useful when one
tries to do (relatively) model-independent inferences or when the
underlying physics is not entirely well established. An example is
the work of \citeN{SNE05}, who found that the polarization plane was
different in the \CaII \, and \HeI \, infrared triplet lines observed
in spicules. The difference can only be understood in terms of Hanle
depolarization, thus providing conclusive model-independent evidence
for magnetic fields in spicules. Another reason could be to obtain
information on the statistical properties of magnetic structures too
small to be resolved and that are expected to exist even with the
resolving power of a 4-m aperture. An example of this approach is the
work of \citeN{SNSA03} who considered 1\arcsec \, resolution
observations of the quiet Sun and showed how the probability
distribution function could be investigated by combining observations
of the 6302 \AA \, and 1.56 $\mu$m pairs of \FeI \, lines.

\begin{figure}
\includegraphics[width=0.45\textwidth]{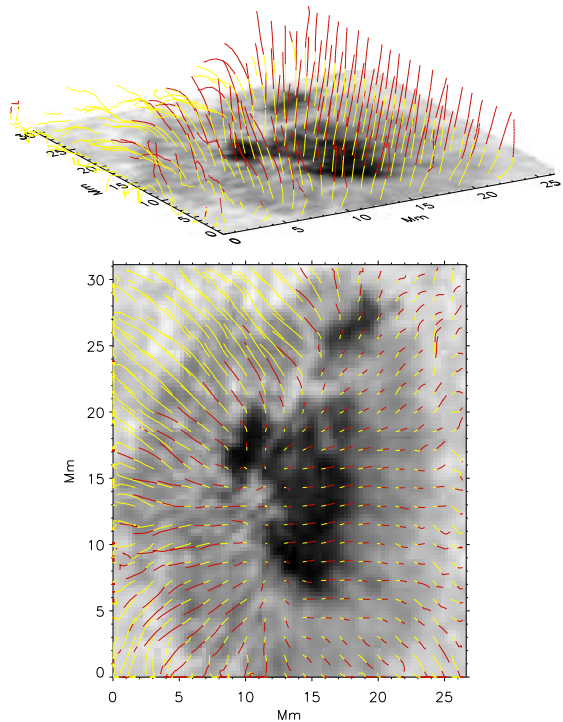}
\caption{Magnetic field lines in a sunspot using \FeI \, and \CaII \,
  lines. Segments in yellow and red represent photospheric
  and chromospheric heights, respectively. }  
\label{spinor}
\end{figure}

Observations of the magnetic field at different heights already exist
but are difficult to make and even more to interpret. Sunspots are a
preferred target (e.g., \citeNP{CCTB06}; \citeNP{SN05c}, see
Fig~\ref{spinor}) with current technology because the signal-to-noise
ratio still represents a challenge in the quiet chromosphere. However,
there have been also breakthroughs in observations of filament loops
(\citeNP{SLW+03}) and network elements (\citeNP{PSNB07}).

\section{Conclusions}

The tremendous leap in spatial resolution that will be provided by the
ATST and the EST will require careful rethinking of most conventional
observing modes, particularly long slit spectroscopy. Without at least
some limited 2D capability, it is not clear how to make these
observations work in practice, particularly with the requirement of
seamless multi-wavelength operation that is so vital to the science
goals of these facilities. It should be noted, though, that a large
aperture is still useful for moderate- or even low-resolution
observations in which one needs to reach very high polarimetric
sensitivities (e.g., for some Hanle effect observations).

Efficient exploitation of the new large solar telescopes with minimum
downtimes will be of paramount importance, especially since it now
appears that they will concentrate most of the resources devoted by
the solar community to observations. This condition implies shifting
the telescope operations to a new model more similar to what is
currently employed in the large night-time facilities or the
space-borne instrumentation such as Hinode. However, there is the risk
that the community will be left without a facility where new
experimental ideas can be tested. It is extremely important that the
new operational model for the ATST/EST considers a provision of at
least a small fraction of the observing time to test new ideas or
generally speaking, high risk/high return potential campaigns.

\acknowledgements 

The author gratefully acknowledges financial support from the Spanish
Ministry of Science through project AYA2007-63881.

\bibliographystyle{../bib/apj}
\bibliography{./articulos,../bib/articulos}

\end{document}